\documentclass[manuscript,showpacs,amsmath,amssymb]{revtex4-1}
\usepackage{amsmath}
\usepackage{amssymb}
\usepackage{graphicx}
\usepackage{dcolumn}
\usepackage{bm}

\begin{document}
\title{Diffusion of Active Particles With Stochastic Torques Modeled
  as $\alpha$-Stable Noise}

\author{J\"org N\"otel$^1$, Igor M. Sokolov$^1$, Lutz
  Schimansky-Geier$^{1,2}$\footnote{alsg@physik.hu-berlin.de}}
\affiliation{$^1$Institute of Physics, Humboldt University at Berlin,
  Newtonstr. 15, D-12489 Berlin, Germany\\$^2$Berlin Bernstein Center
  for Computational Neuorscience, Philippstra\ss e 12, D-10115 Berlin,
  Germany}

\begin{abstract}
  We investigate the stochastic dynamics of an active particle moving at a constant
  speed under the influence of a fluctuating torque. In our model the angular velocity is generated by a
  constant torque and random fluctuations described as a L\'evy-stable noise. 
  Two situations are investigated. First, we study white
  L\'evy noise where the constant speed and the angular noise generate
  a persistent motion characterized by the persistence time $\tau_D$.
  At this time scale the crossover from ballistic to normal diffusive
  behavior is observed. The corresponding diffusion
  coefficient can be obtained analytically for the whole class of symmetric
  $\alpha$-stable noises. As typical for models with noise-driven angular
  dynamics, the diffusion coefficient depends non-monotonously on the
  angular noise intensity. As second example, we study angular noise as described by
  an Ornstein-Uhlenbeck process with correlation time $\tau_c$ driven
  by the Cauchy white noise.  We discuss
  the asymptotic diffusive properties of this model and obtain the
  same analytical expression for the diffusion coefficient as in the
  first case which is thus independent on $\tau_c$. Remarkably, for $\tau_c>\tau_D$ the 
  crossover from a non-Gaussian to a
  Gaussian distribution of displacements takes place at a time $\tau_G$ which can be 
  considerably larger than the persistence time $\tau_D$. 
\end{abstract}
\maketitle

\section{Introduction}
Over the past few years an increasing interest to the motion of living organisms of different 
size and complexity (movement ecology \cite{Natan}) has lead to a large number of new experiments, 
some of them based on quite elaborated tools, allowing for study of this motion and for description 
of its observed trajectories. Examples are the works on dicstyostelium \cite{Bodecker}, also 
under influence of chemotactic stimulus \cite{Anselem}, on human motile keratinocytes and 
fibroblasts \cite{Selmeczi}, on motile pieces of physarum \cite{Rodiek}, 
on flagellate eukaryotes (Euglena) \cite{Euglena}, also in time dependent
light fields \cite{Romensky}, and on more complex organisms as gastropods \cite{Seuront}, 
zooplankton \cite{Ordemann}, birds \cite{Edwards} and zebra-tail fish \cite{Gautrais}. 
One has also considered motion in the presence of boundaries which might substantially 
change its effective properties \cite{TefLow08,TefLo09}. Similar conceptual approaches were
used to describe the motion of non-living motile objects, so called self-propelled
particles, exhibiting similar trajectories \cite{He,Takagi,Hagen}.

Theoretical modeling of self-motile objects goes back to the beginning of
the last century \cite{Pearson,Fuerth}. Trajectories often appear to be
stochastic \cite{Berg,Campos,Romanczuk}, thus special attention has been
paid to the development of different stochastic models for various
physical or biological situations. These investigations yielded a great variety of possible
mathematical models for the description of motion of self-propelled objects. Examples are the discrete
hopping with a given turning angle distribution \cite{lsgzigzag,HaLsg08}, 
the run and tumble model \cite{Thiel} with polynomial waiting time densities exhibiting
superdiffusive behavior etc. In order to determine the
respective mean squared displacement (MSD) of the diffusive motion 
Langevin equations with associated kinetic equations for the probability densities have been
introduced to describe propulsive motion under influence of noise sources, as modeled by white
Gaussian noise \cite{Mikhailov,Blum,Peruani}, by Gaussian Ornstein
Uhlenbeck process \cite{Motsch,Weber}, or by dichotomic Markovian process \cite{Weber_Sok}.
Escape rates of active particles \cite{Burada}, of active particles in
external fields \cite{Ebeling,Geiseler} and active transport in
cells \cite{Godec} have been studied as well. Crawling patterns of
Drosophila larvae motion where analyzed in \cite{Gunther} by using a bimodal persistent random walk model.  
Here, the four parameters of the corresponding model were proved to be distinct for larvae with specific genetic
mutations. This work elucidates the usefulness of elaborate, multiparametric models.

The origin of stochasticity, i.e. the nature of the noise sources, might be
quite different, e.g. they may be due to neuronal activity, or due to interactions with the environment or with neighbors. 
The patchiness of the distribution of prey \cite{Edwards} can be also
considered as a noise source.  There is still a debate on how to correctly
describe various patterns of animal motion \cite{Reynolds}. 
In the present work we extend the model of an active particle moving
at a constant speed with a stochastic angular dynamics \cite{Weber} to a more general case, 
when the random torques are described by an $\alpha$-stable noise. These L\'evy or $\alpha$-stable
noise sources lead to a directed motion interrupted by large jumps in
the angle of orientation \cite{Dybiec}. The sudden
changes in the direction of movement and also the longer periods of
curling appearing in this model might remind in particular the run and tumble processes \cite{Thiel}.

In what follows we discuss two different angular dynamics. In the first case, Sec. \ref{sec_tor_white}, we
consider the motion in the presence of a constant torque and of a source of a symmetric $\alpha$-stable 
white noise. Afterwards, in Sec. \ref{Sec_OUP}, we look at a non-white noise as generated by an Ornstein-Uhlenbeck
process (OUP) driven by a Cauchy noise source. For both dynamics we
derive the analytical expressions for the mean
squared displacement (MSD) and for the effective diffusion coefficient, and
discuss the crossover from the ballistic motion to normal diffusion in coordinate space. 
We show that the MSD, the diffusion coefficient, and the crossover time
do not depend on the correlations in the noise source.

Recently, experiments on beads diffusing on macromolecules were performed \cite{Wang,Wangnat} and caught attention.
In these experiments single particle tracking was used to follow the trajectories 
of the beads. The beads performed (passive) Fickian
diffusion, but a non-Gaussian distribution of displacements was observed.
These observations point out the necessity to
pay special attention to the moments of the displacement distributions higher than the MSD.
In our model indeed, at difference to the behavior of the MSD, the higher moments of the displacement 
are sensitive to correlations.
In particular, in the diffusive regime of the OUP-driven active particles we find an exponential distribution
of displacements at short time lags while at longer time lags this distribution tends to a Gaussian,
an observation similar to the one reported by Wang et al. \cite{Wang,Wangnat}.
In our case this effect is caused by correlations in the noise which create an additional persistence in the motion. 
The crossover time after which the Gaussian distribution is established
depends strongly on these correlations.

\section{Torque and white $\alpha$-stable angular noise}
\label{sec_tor_white}
The the time evolution of the position
vector $\vec{r}=\left(x(t),y(t)\right)$ of the particle is described by the following set of equations:
\begin{eqnarray} 
\frac{\text{d}\vec{r}}{\text{d}t}=v_0
\begin{pmatrix}
         \cos\varphi(t) \\
         \sin\varphi(t)
        \end{pmatrix}
\label{v_dot}
\end{eqnarray}
\begin{eqnarray} 
\frac{\text{d}\varphi}{\text{d}t}=\Omega+\frac{\sigma}{v_0}\xi(t) 
\end{eqnarray}
where the angle $\varphi(t)$ gives the orientation of the velocity
vector whose absolute value is $v_0$. $\Omega$ is a constant torque. The noise $\xi(t)$ considered
here is a $\alpha$-stable noise.  According to
Ditlevsen \cite{Ditlevsen} and Schertzer \cite{Schertzer}, the associated
Fokker-Planck-Equation for such a system is given by:
\begin{eqnarray} 
\frac{\partial}{\partial t}P(\varphi,t)=-\Omega \frac{\partial}{\partial \varphi} P +\frac{\sigma^\alpha}{v_0^\alpha}\frac{\partial^\alpha}{\partial|\varphi|^\alpha}P(\varphi,t).
\label{fpe}
\end{eqnarray}
Herein
\begin{eqnarray} 
\frac{\partial^\alpha}{\partial|\varphi|^\alpha}P(\varphi)=-\frac{1}{2\pi}\int_{-\infty}^{\infty}dke^{-ik\varphi}|k|^\alpha P(k)
\end{eqnarray}
stands for the $\alpha$-th symmetric Riesz-Weyl fractional derivative. The case $\alpha=2$
corresponds to a Gaussian white noise. The parameter
$\alpha$ controls the sudden large changes in the velocity orientation. 
For lower values of $\alpha$ the noise distribution becomes sharper around the center and the tails
become more pronounced, so that the probability of large sudden changes in $\varphi$ increases, while small changes become less likely. 

The conditional probability density of the orientation $\varphi$ at time $t$, given the initial angle $\varphi_0$ at time $t_0$  
of the velocity vector is given by the following expression \cite{Schertzer}:
\begin{eqnarray} 
P(\varphi,t|\varphi_0,t_0)=\frac{1}{2\pi}\int_{-\infty}^{\infty}dke^{-ik(\varphi-\varphi_0)}e^{ik\Omega (t-t_0)}e^{-\left(\frac{\sigma}{v_0}|k|\right)^\alpha (t-t_0)} .
\label{fok_white_phi}
\end{eqnarray}
In the following we will set the initial values to $t_0=0$ and $\varphi_0=0$.
With the help of Eq.\eqref{fok_white_phi} the MSD for
this dynamics can be easily calculated using the Green-Kubo relation.
For simplicity of notation, we  move the origin of the coordinate system to the initial 
position of the particle at time $t_0$. Hence we have as as initial values $\vec{r}_0=0$ or $x_0=x(t_0)=0$ and $y_0=y(t_0)=0$, 
if not explicitly defined otherwise. We assign $r(t)=|\vec{r}|= \sqrt{x^2(t)+y^2(t)}$. 
The MSD becomes:
\begin{eqnarray} 
\langle r^2(t)\rangle =2v_0^2t\int_0^t d\tau
\left(1-\frac{\tau}{t}\right)\cos(\Omega
\tau)e^{-\left(\frac{\sigma}{v_0}\right)^{\alpha}\tau}.
\label{anoise}
\end{eqnarray}
Introducing $\gamma=\left(\frac{\sigma}{v_0}\right)^{\alpha}$ and
$a=\Omega^2+\gamma^2$ leads to the following form of the  MSD:
\begin{eqnarray} 
\langle r^2(t)\rangle =2v_0^2\left(\frac{\gamma
  t}{a}+\frac{\left(\gamma^2-\Omega^2\right)\left(\cos(\Omega t)
  e^{-\gamma t} -1\right)-2\gamma\Omega \sin(\Omega t)e^{-\gamma
    t}}{a^2} \right).
\label{msd}    
\end{eqnarray}
For short times $t$ the MSD shows ballistic behavior, i.e. $\langle r^2(t)\rangle=v_0^2t^2$. 
For times $t\gg\tau_D$ with
\begin{eqnarray} 
\tau_D=\left(\frac{v_0}{\sigma}\right)^\alpha
\label{taud}
\end{eqnarray}
being the crossover time, the motion becomes diffusive.  The persistence length of the motion is 
\begin{eqnarray}
l_D=v_0\tau_D=v_0\left(\frac{v_0}{\sigma}\right)^\alpha
\label{persi}
\end{eqnarray}
and by that dependent on the noise type through the parameter
$\alpha$.  The effective diffusion coefficient $D_{\text{eff}}$ is given by 
\begin{eqnarray} 
D_{\text{eff}}=\lim\limits_{t\to\infty}\frac{\langle r^2(t)\rangle}{4t}=v_0^2\frac{\left(\frac{\sigma}{v_0
  }\right)^\alpha}{2\left(\Omega^2+\left(\frac{\sigma}{v_0
  }\right)^{2\alpha}\right)} .
\label{diff}
\end{eqnarray}
\begin{figure}[ht]
\includegraphics[width=.65\textwidth]{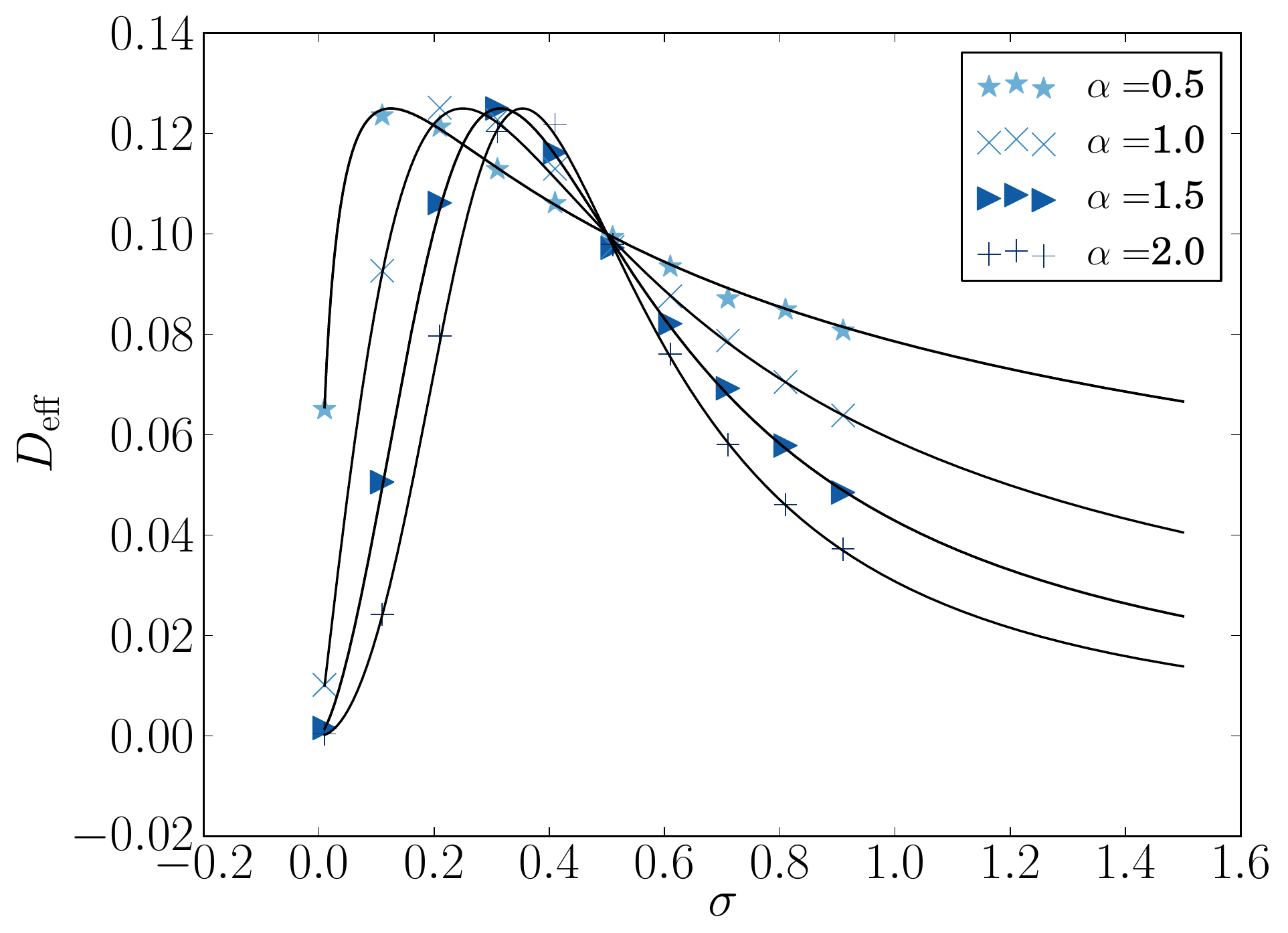}
   \caption{\small The effective diffusion coefficient $D_{\text{eff}}$ as a function of the parameters of the noise: Simulation (symbols) and theory (curves) for
    $\Omega=0.5$. }
\label{msd_anf}
\end{figure}
In the case of Gaussian white noise $(\alpha=2)$ the result of Weber
et al.\cite{Weber} is recovered. For $\Omega=0$ the effective
diffusion coefficient $D_{\text{eff}}=v_0^{2+\alpha}/(2\sigma^\alpha)$
decays with the growth of the scale parameter $\sigma$ as a power law, with a power given by the index of the L\'evy distribution. 
Using Eq.\eqref{persi} we see that the effective diffusion coefficients becomes
$D_{\text{eff}}=v_0 l_D/2$.

In Fig.\ref{msd_anf} we show the effective diffusion coefficient $D_{\text{eff}}$ for various values of $\alpha$ and
non-zero torque, as obtained in simulations and from (Eq.\eqref{diff}). The effective
diffusion coefficient always shows a maximum at $\sigma_{\text{max}}=v\Omega^{1/\alpha}$, where it is equal to
$D_{\text{eff}}^{\text{max}}=v_0^2/(4\Omega)$. This maximal value of the diffusion coefficient does not depend on the 
properties of the noise. For $\Omega\ne0$ the
diffusion coefficient vanishes for $\sigma\rightarrow0$ and for
$\sigma\rightarrow\infty$.  How fast it vanishes depends on
$\alpha$. The higher the value of $\alpha$ the smaller becomes the width of the peak.
All curves intersect at $\sigma=v_0$. One also notices that the diffusion for
$\alpha<2$ for a given $\sigma$ can be both faster or slower than in the Gaussian case. 
For small noise intensities the motion of the
particle is dominated by the torque, the particle moves in circles and
the diffusion is slow. Increasing the noise stretches the trajectories
till the diffusion reaches the maximum. Behind the maximum the motion
becomes noise-dominated, and the diffusion coefficient falls again.

Simulations show that for times $t$ larger than the crossover time $t\gg \tau_D$ the transition probability density of the spatial process
becomes Gaussian,
\begin{eqnarray} 
\label{eq:normal}
  P(x,y,t|x_0,y_0, t_0)=\frac{1}{4\pi D_{\text{eff}}(t-t_0)}\exp\left(-\frac{(x(t)-x(t_0))^2+(y(t)-y(t_0))^2}{4D_{\text{eff}}(t-t_0)}\right).
\end{eqnarray}
Notably, as simulations show, the crossover time between ballistic and
diffusive behavior coincides with the establishment of this Gaussian
displacement distribution.
The distribution of the final position depends only on the distance between the final and the initial points, 
and the dependence on the angle of the vector $\vec{r}(t)$ with respect to the initial orientation of the velocity is lost. 
As we had moved the origin of the coordinate system to the position of the particle at time $t_0$ ($x(t_0)=y(t_0)=0$), we write $r(r)$ 
instead of $\Delta r(t)$ for the absolute displacement. We set $t_0=0$.  The statistics of the absolute displacements is given by the Rayleigh
distribution. The probability density of $r(t)$ reads:
\begin{eqnarray}
  \label{eq:Rayleigh}
  P(r,t)= \frac{r}{2 D_{\text{eff}}t} \exp\left( - \frac{r^2}{4D_{\text{eff}}t}\right)
\end{eqnarray}
where we omit $t_0=0$. 

\section{Ornstein-Uhlenbeck process for the Cauchy distribution} 
\label{Sec_OUP}
We now consider colored noise in the angular dynamics. We use the
Ornstein-Uhlenbeck process (OUP) with a Cauchy noise ($\alpha=1$) for the torque. 
The equations for the spatial coordinates and for the time evolution of the velocity (Eq.\eqref{v_dot}) 
remain the same as before. 
The angular dynamics is given by
\begin{eqnarray}
\frac{\text{d}\varphi}{\text{d}t}&=&\Omega +\frac{1}{v_0}\theta(t)\\
\frac{\text{d}\theta}{\text{d}t}&=&-\frac{1}{\tau_c}\theta(t)+\frac{\sigma}{\tau_c}\xi(t),
\label{OUP}
\end{eqnarray}
where $\xi(t)$ is white noise with increments following the Cauchy distribution.
In the limit $\tau_c\rightarrow0$  in Eq.\eqref{OUP} this model coincides with the model from the previous section for $\alpha=1$. 
In the limit $\tau_c\rightarrow\infty$ the noise vanishes, but if one takes $\sigma \propto \tau_c$ another interesting limiting
case emerges: now the angle $\theta(t)$ becomes a L\'evy process: Since in this limit $\dot{\theta}=\xi(t)$ we get $\theta(t)=\int_0^t\xi(t')dt'=L(t)$
(here we took $\sigma = \tau_c$).

By introducing the OUP, we get a stronger persistence in the particle's motion and introduce a new time
scale, the correlation time $\tau_c$. For times larger the crossover 
time $\tau_D$ the particles again perform a diffusive motion. Surprisingly, 
as we proceed to show, the expressions for the crossover time, the MSD and 
the diffusion coefficient coincide with Eqs.\eqref{msd} \eqref{taud} and \eqref{diff} of the previous section. 
The onset of the asymptotic Gaussian displacement distribution is however characterized by a new time scale $\tau_G$
which is in general different from $\tau_D$ as given by Eq.\eqref{taud}.

\subsection{Mean squared displacement and effective diffusion coefficient } 
The Fokker-Planck equation the OUP with Cauchy noise, Eq.(\ref{OUP}), has the following form:
\begin{eqnarray}
\frac{\partial}{\partial t}P(\theta,t)=\left(\frac{\partial}{\partial \theta}\frac{\theta}{\tau_c}+\frac{\sigma}{\tau_c}\frac{\partial}{\partial |\theta|}\right)P(\theta,t).
\label{fok_theta}
\end{eqnarray}
The transition probability for Eq.(\ref{fok_theta}), given the initial condition $\theta_0$ at time $t_0$, was first obtained in Ref. \cite{West}:
\begin{equation}
P(\theta,t|\theta_0,t_0)=\frac{1}{\pi}\frac{\sigma(1-e^{-(t-t_0)/\tau_c})}{(\theta-\theta_0e^{-(t-t_0)/\tau_c})^2+\sigma^2(1-e^{-(t-t_0)/\tau_c})^2}.
\label{Ptheta}
\end{equation}
The width of the stationary Cauchy distribution for $\theta$ depends on $\sigma$ only. 
For $\tau_c\rightarrow0$ Eq.(\ref{fok_phi}) becomes the white noise limit (Eq.(\ref{fok_white_phi})). 
 The angular transition probability density was determined in Ref.\cite{Garbaczewski}:  
\begin{equation}
P(\varphi,t|\varphi_0,t_0)=\frac{a}{\pi}\frac{1}{(\varphi-\varphi_0 -\theta_0\frac{\tau_c}{v_0}(e^{-t_0/\tau_c}-e^{-t/\tau_c}))^2+a^2(t,t_0)}
\label{fok_phi_garb}
\end{equation}
With $a=\frac{\sigma}{v_0} \left(t-t_0-\tau_c(e^{-t_0/\tau_c}-e^{-t/\tau_c} ) \right)$ and $\theta_0$, $\varphi_0$ being the values of $\theta$ and $\varphi$ at time $t_0$.
The angular variable $\varphi$ is a deterministic functional (integral) of the stochastic variable $\theta$, and in \cite{Garbaczewski} it is pointed out
that for fixed initial $\theta_0$ Eq.\eqref{fok_phi_garb} represents a Markov process. The Cauchy case is the only case of an  OUP with an $\alpha$-stable noise where the 
integrated process is Markovian, and therefore the simplest one for the further analytical treatment. 
Adding the torque and setting the initial time $t_0$ to zero results in   
 \begin{equation}
P(\varphi,t|\varphi_0,0)=\frac{a}{\pi}\frac{1}{(\varphi-\varphi_0-\Omega t -\theta_0\frac{\tau_c}{v_0}(1-e^{-t/\tau_c}))^2+a^2(t,0)}.
\label{fok_phi}
\end{equation}
Using Eq.\eqref{fok_phi} one easily derives the expression for the MSD. 
Since the MSD does not depend on the initial angle, we set $\varphi_0=0$. 
The second initial value $\theta_0$ has been averaged over the stationary 
limit of Eq.(\ref{Ptheta}) which is achieved for $t\rightarrow\infty$. 
The MSD for the OUP-Cauchy is equal to that of an active particle from the previous
Sec. \ref{sec_tor_white}, Eqs.\eqref{anoise} and \eqref{msd} with $\alpha=1$.  Averaging over initial
conditions cancels any additional time dependence which would reflect
the correlation behind the evolution of the direction of motion, in contrast to the OUP with Gaussian white
noise \cite{Uhlenbeck,Weber}. 

\begin{figure}[ht]
  \includegraphics[width=.65\textwidth]{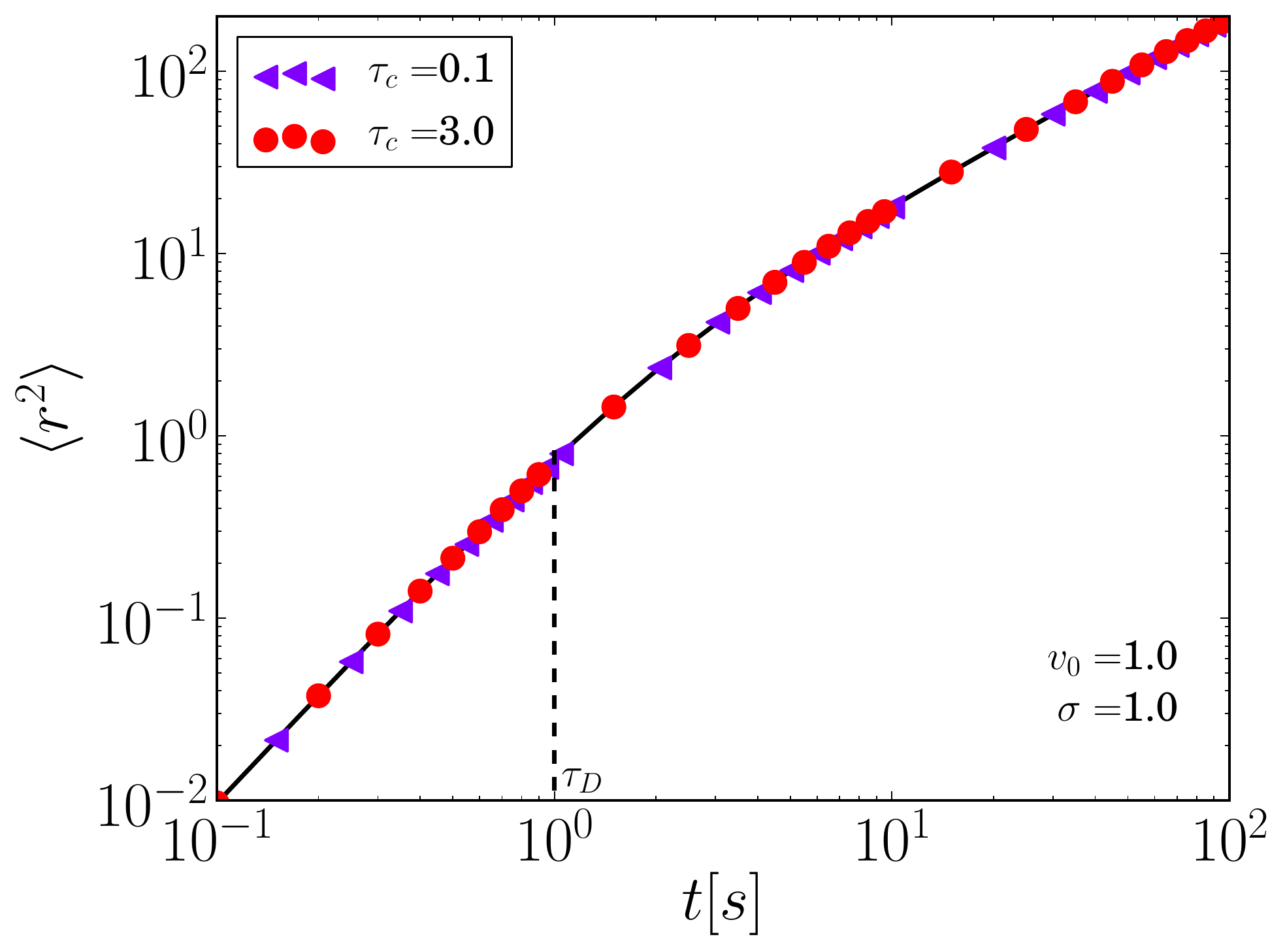}

   \caption{\small  MSD from simulations (colored symbols) for two different correlation times $\tau_c$, $\sigma=v_0=1.0$, $\Omega=0$, 
   black line corresponds to Eq.\eqref{msd} with $\alpha=1$, theory and simulation align perfectly, the MSD does not depend on the correlation time $\tau_c$
   the crossover time from ballistic to diffusive motion 
   is indicated by dashed line at time $\tau_D$.}
   \label{figoupmsd}
\end{figure}

As the MSD in our correlated case is the same as for an active particle with white angular noise with $\alpha=1$, 
the following properties stay the same: For times $t<\tau_D$ with $\tau_D=1/\gamma$ the active particle moves ballistic.  
For times  $t\gg\tau_D$ the regime of normal or Fickian diffusion sets on, with the diffusion coefficient
given by Eq.\eqref{diff} with $\alpha=1$. The crossover from ballistic to diffusive behavior
is well seen in Fig.\ref{figoupmsd}. The behavior of the effective
diffusion coefficient in dependence on other parameters was already shown in Fig.\ref{msd_anf} for
$\alpha=1$. For non-vanishing torque this diffusion coefficient has a maximum at $\sigma_{\mathrm{max}}$ and tends to zero
for large and for small noise intensities.

\subsection{Distribution of displacements}
It is remarkable that the MSD and the diffusion coefficient are
independent of the correlation time $\tau_c$, so that the underlying correlations cannot be detected from measuring the
MSD. Therefore one might expect to observe a Gaussian distribution for the
displacements, Eq.\eqref{eq:normal}, for times $t > t_D$ and, respectively, the Rayleigh
distribution, Eq.\eqref{eq:Rayleigh}, for the absolute displacement for all such times.
This expectation is however wrong.
As we show in the following plots, the probability density of the absolute displacement
 $P(r,t)$ deviates strongly from the Rayleigh distribution
even for times well beyond the crossover time to diffusive behavior
$\tau_D$. Depending on $\tau_c$, the asymptotic convergence to the expected Gaussian and Rayleigh distributions 
which are independent of $\tau_c$ may take place only at times which are considerably larger.

\begin{figure}[ht]
  \includegraphics[width=.49\textwidth]{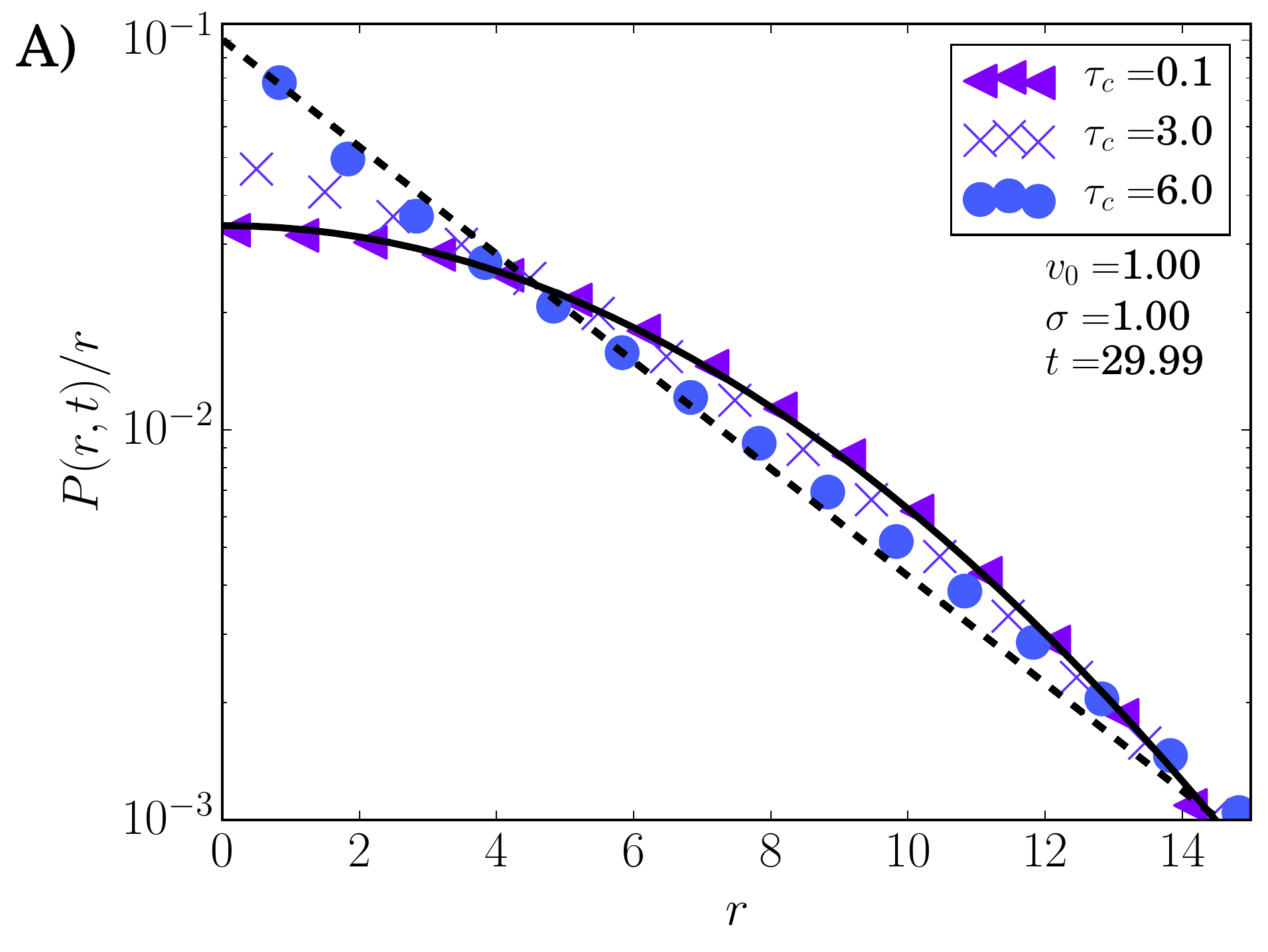}
   \includegraphics[width=.49\textwidth]{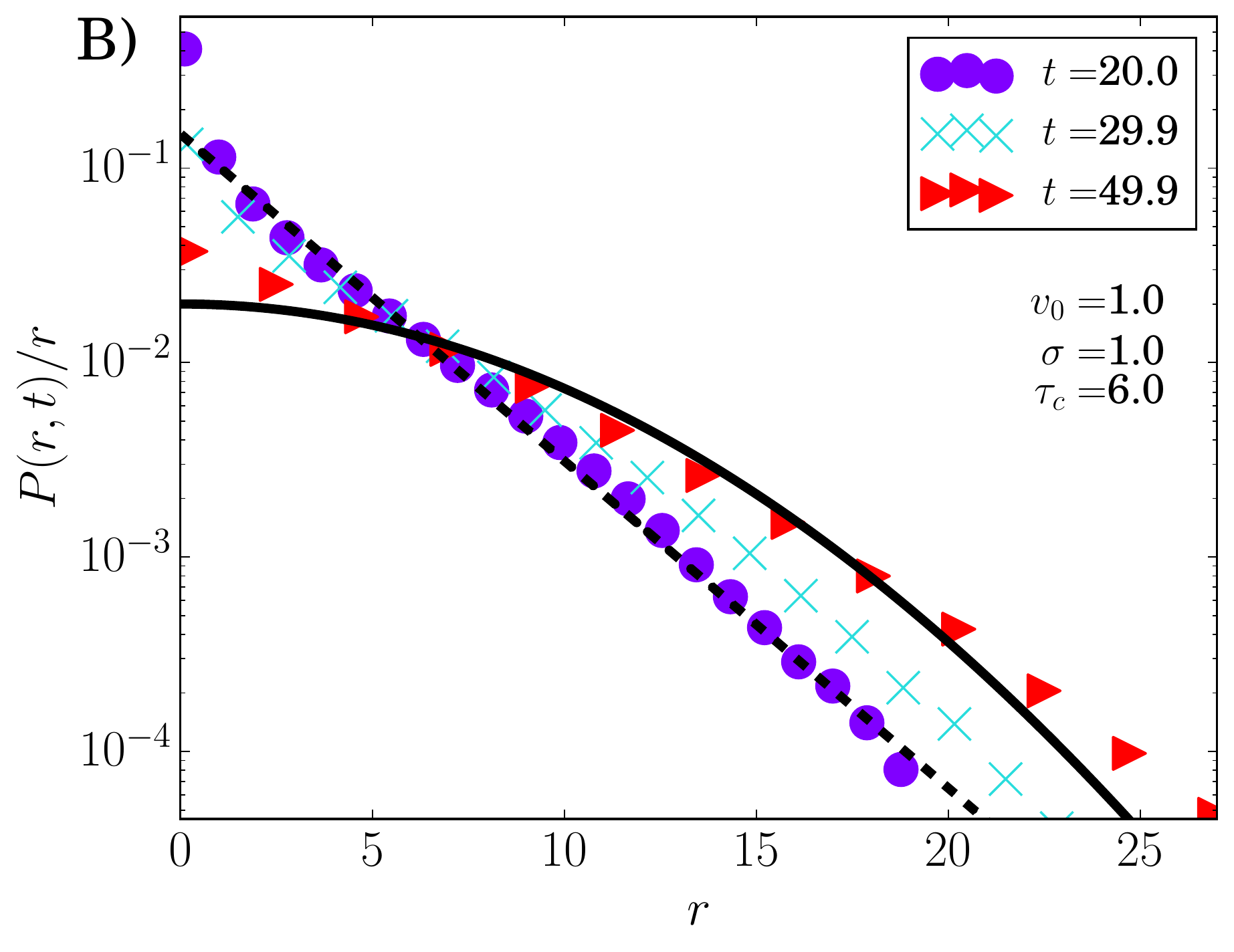} 
   \caption{\small A): log plot of simulation results (colored) for the displacement distributions scaled by
     $r$ at $t= 29.9\gg \tau_D$ for given correlation time $\tau_c$.
     Black line corresponds to the absolute displacement distribution
     Eq.\eqref{eq:Rayleigh} at the given time $t$, with $D_\text{eff}$ from
     Eq.(\ref{diff}),($\alpha=1$). The black dashed
     line shows the exponential distribution from
     Eq.\eqref{eqn_displ_exp_f} with $D_\text{eff}$ from
     Eq.\eqref{diff}, with $\alpha=1$ at the given time.; B): The logarithmic plot of the time evolution displacement
     distributions rescaled by $r$ for indicated times and
     parameters. Colored symbols correspond to simulations. The black line corresponds to 
     Eq.\eqref{eq:Rayleigh} for ($t=49.9$). The black dashed
     line shows the exponential distribution from
     Eq.\eqref{eqn_displ_exp_f} with $D_\text{eff}$ from
     Eq.\eqref{diff},($\alpha=1$) at the given time. At time $t=20.0s$ the displacement distribution
     (blue circles) displays an exponential behavior and at time $t=49.9s$ (red triangles) it has come closer to a Gaussian.}
   \label{figspat}
\end{figure}

In Fig. \ref{figspat}A the absolute displacement distributions rescaled by $r$
for three different correlation times $\tau_c$ at time $t=29.9$ are shown. The
chosen time is well beyond the crossover time $\tau_D$ to the
diffusion regime, i.e. $t\gg\tau_D$, and also much larger than
correlation time, $t\gg\tau_c$. All three distributions are different.
Only for the shortest correlation time the simulation results and the prediction of Eq.\eqref{eq:Rayleigh} (the black line in Fig. \ref{figspat}A) 
coincide. For longer correlation times the absolute displacement
shows an exponential behavior (black dashed line, Eq.\eqref{eqn_displ_exp_f}). 
Such behavior is not observed in simulations of particles driven by an OUP with Gaussian white noise.  
In Fig. \ref{figspat}B the time evolution of the absolute displacement
is presented. The results for three different time lags, all within the
diffusion regime, are plotted. The black line corresponds to the
result of Eq.\eqref{eq:Rayleigh} at time lag t=49.9, with the
effective diffusion coefficient according to Eq.\eqref{diff} with $\alpha=1$.  As the
simulations show, the distribution (red triangles in
Fig.(\ref{figspat}B)) at this time has still not fully
approached its asymptotic shape (black line).  At smaller time lags the displacements
clearly deviate from this line, exhibiting an exponential behavior. The black dashed
line shows such an exponential decay.  

\begin{figure}[ht]
   \includegraphics[width=.49\textwidth]{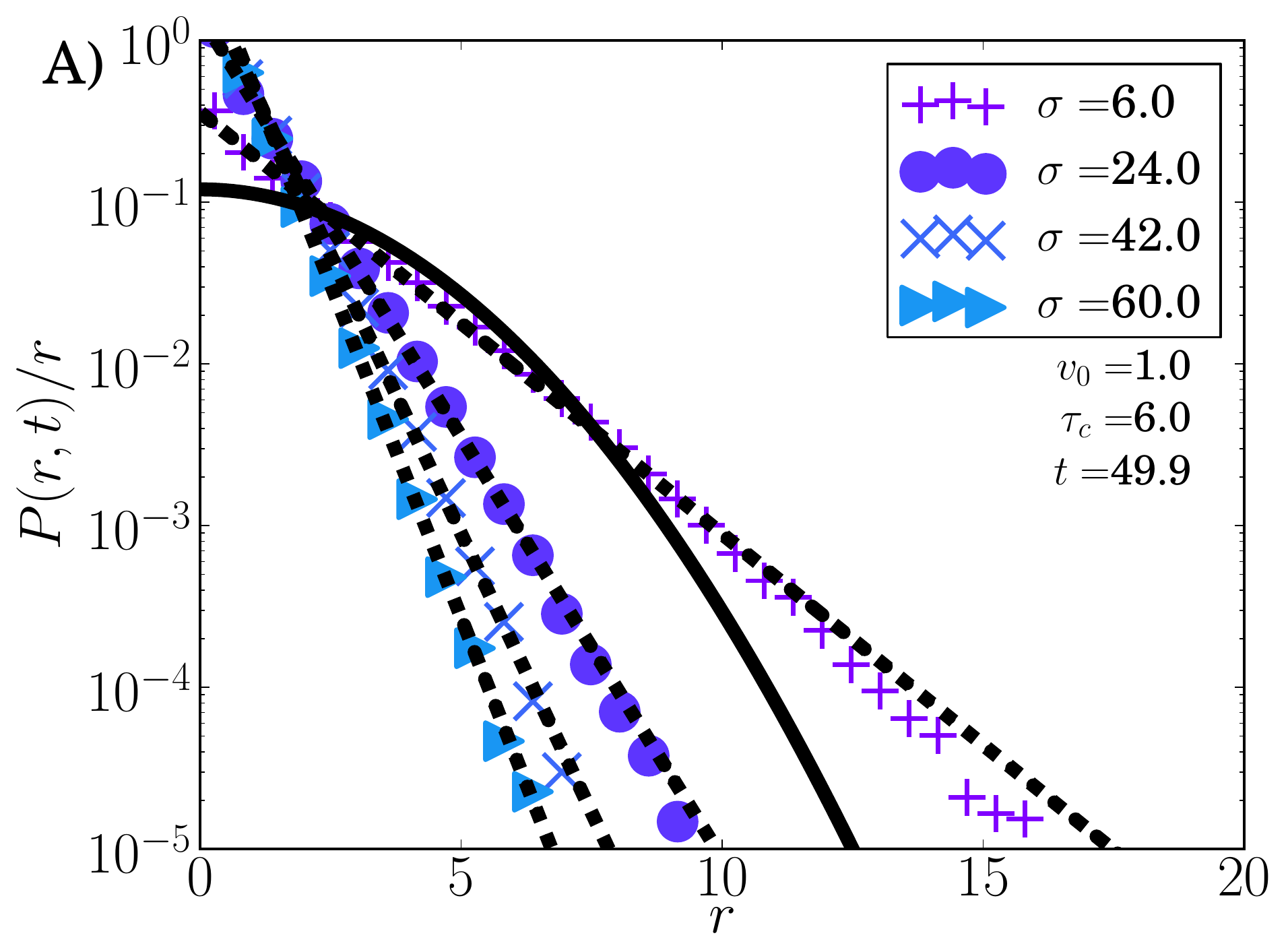} 
   \includegraphics[width=.49\textwidth]{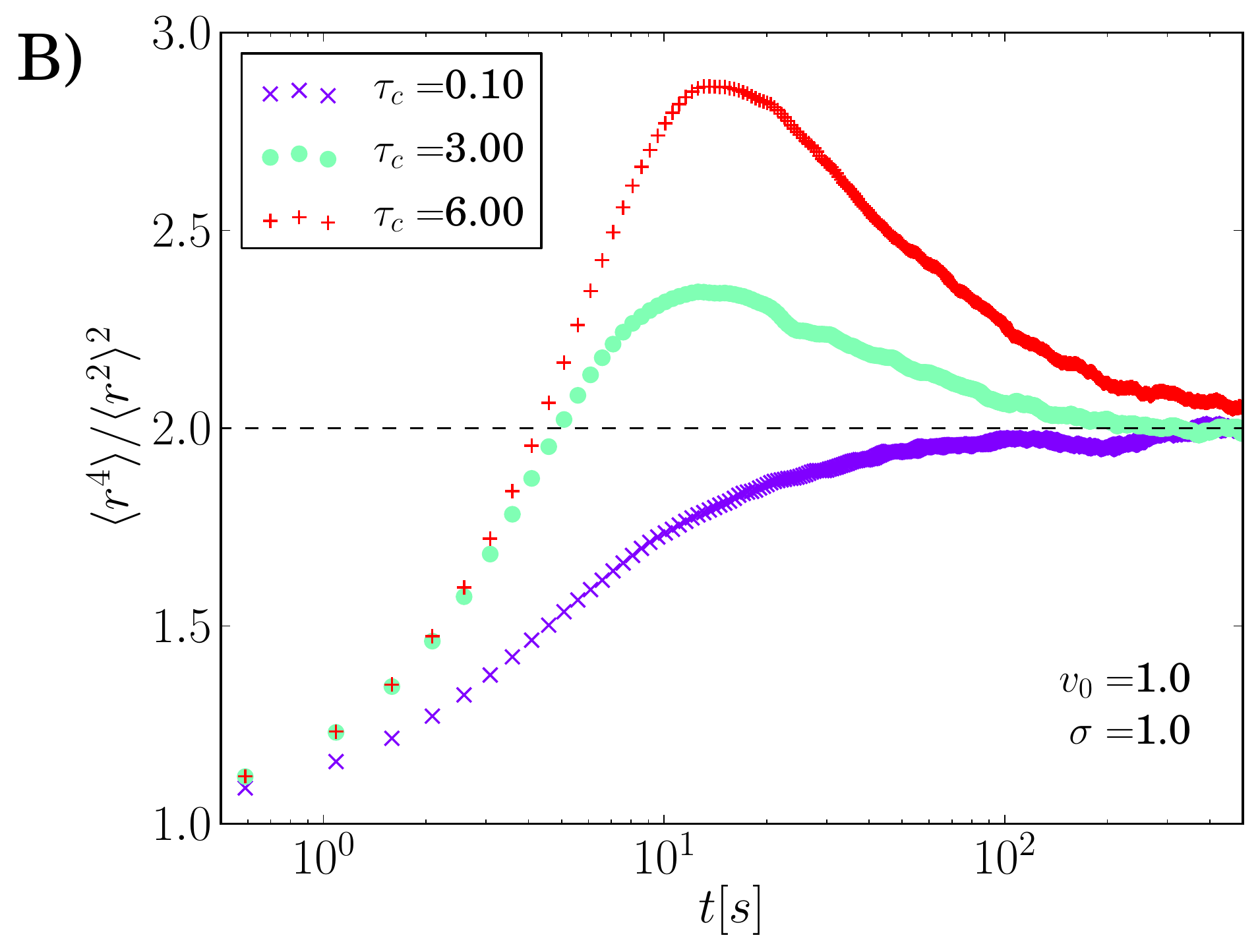} 
   \caption{\small A): Probability densities of displacement distributions scaled by
     $r$ for various intensities of the Cauchy distributed noise
     in Eq.\eqref{OUP}. Colored symbols correspond to simulations. Black line
     is the distribution from Eq.\eqref{eq:Rayleigh}. Black
     dashed lines are exponentials from
     Eq.\eqref{eqn_displ_exp_f}. B): Simulation results for the kurtosis of the distributions for
     different correlation times. The values approach asymptotically
     the Gaussian limit (dashed black line) indicating the crossover
     to a Gaussian displacement distribution.}
   \label{figth}
\end{figure}

Fig. \ref{figth}A shows the the absolute displacement distribution for different
values of the noise strength $\sigma$. Colored symbols stand for simulations. The
black line shows the Rayleigh distribution from Eq.\eqref{eq:Rayleigh} 
rescaled by the distance $r(t)$. The black dashed lines are exponentials 
from Eq.\eqref{eqn_displ_exp_f}). For
higher noise intensities the distributions decay faster.
Higher noise intensities correspond to an increase in directional
changes which then lead to smaller spatial increments in a given time
interval. 

It is in principle possible (but tedious) to calculate higher moments of the displacement using Eq.\eqref{fok_phi}.
We choose here a phenomenological approach. We make the following exponential ansatz 
for the normalized distribution density of the displacements with the time dependent characteristic length scale $l(t)$:
\begin{eqnarray}
 P(x,y,t|x_0=0,y_0=0,t_0=0)= \frac{1}{2\pi l^2(t)}\exp\left(-\frac{ \sqrt{x^2+y^2}}{l(t)}\right)
\end{eqnarray}
In agreement with our initial condition, this scale at initial time $l(t_0)$ should vanish, i.e. $l(t_0)=0$. For the absolute
displacement the probability distribution reads:  
\begin{eqnarray}
 P(r,t)= \frac{r}{l^2(t)}\exp\left(-\frac{r}{l(t)}\right)
 \label{eqn_displ_exp}
\end{eqnarray}
We require that the expectation value $\langle r^2(t)\rangle$ for the squared absolute 
displacement (Eq.(\ref{eqn_displ_exp})) is equal to the long time limit of the
calculated MSD (Eq.(\ref{msd}))
\begin{eqnarray}
\langle r^2(t)\rangle=4D_\text{eff}t.  
\end{eqnarray}
On the other hand, it results from Eq.(\ref{eqn_displ_exp}) that
$\langle r^2(t)\rangle=6l^2(t)$. Therefore, the absolute displacement distribution
$P(r,t)$ reads:
\begin{eqnarray}
P(r,t)=\frac{3r}{2D_\text{eff} t} \exp\left(-\frac{r}{\sqrt{\frac{2}{3}D_\text{eff} t}}\right)
\label{eqn_displ_exp_f}
\end{eqnarray}
The black dashed lines in Fig.(\ref{figspat}) and Fig.(\ref{figth}A) correspond to
Eq.(\ref{eqn_displ_exp_f}). This distribution fits the
simulations very well for times where the displacement is not yet Gaussian,
although Eq.({\ref{eqn_displ_exp_f}}) does not depend on the correlation time. 

Fig. \ref{figth}B shows simulations for the fourth moment
of the displacement divided by the squared MSD (kurtosis). 
In two dimensions a value of two corresponds to a Gaussian distribution, shown by the horizontal black dashed line in Fig.(\ref{figth}B).
A higher value of the kurtosis 
indicates heavier tails and sharper central peak in the distribution, 
while a smaller value corresponds to lighter tails and flatter peak. 
In experiments it can be calculated from the particle positions at different instances of time.

In Fig. \ref{figth}B the kurtosis for three different correlation times $\tau_c$ is plotted.
With growing time asymptotically the Gaussian limit is reached for all correlation times.
For short correlation times $\tau_c \to 0$ the limit is reached from below, while for larger $\tau_c$ 
the kurtosis exhibits a maximum and then approaches the limit from above, indicating 
the observed transient exponential or heavy-tailed regime. The height of the maximum grows 
with increasing correlation time, so the deviation from the Gaussian displacement becomes more 
pronounced in the transient regime. With increasing $\tau_c$ the time till the asymptotics is 
reached also increases.

We give here an estimate of the crossover time $\tau_G$ from non-Gaussian to
Gaussian displacement distributions. We require, for $t>\tau_D$, that the MSD equals the squared correlation 
length $l_c=v_0\tau_c$, a typical length of the trajectory within the correlation time $t_c$ of the noise, i.e.
\begin{eqnarray}
\langle r^2(t=\tau_G)\rangle= l_c^2.
\end{eqnarray}
Note that the length $l_c$ differs from the persistence length, as introduced in Eq.\eqref{persi},
being the characteristic scale of transition from ballistic to diffusive motion.
Since the particle is already in the diffusive regime the MSD becomes
$4D_\text{eff} \tau_G$ and in consequence we get:
\begin{eqnarray}
\tau_G = \frac{v_0^2\tau_c^2}{4D_\text{eff}} \ge \tau_D
\end{eqnarray}
For zero torque $\Omega=0$ in Eq.(\ref{diff}) $\alpha=1$ the estimate for the crossover time becomes:
\begin{eqnarray}
\tau_G=\frac{\tau_c^2}{2\tau_D}
\end{eqnarray}
Comparing this result for $\tau_G$ with Fig. \ref{figspat}B indicates
that this time $\tau_G$ is a lower bound for the establishment of the Rayleigh
displacement distribution from Eq.\eqref{eq:Rayleigh}. Hence, for $t\gg\tau_G$ the  distribution of displacements
becomes Gaussian.

\begin{figure}[ht]
  \includegraphics[width=.49\textwidth]{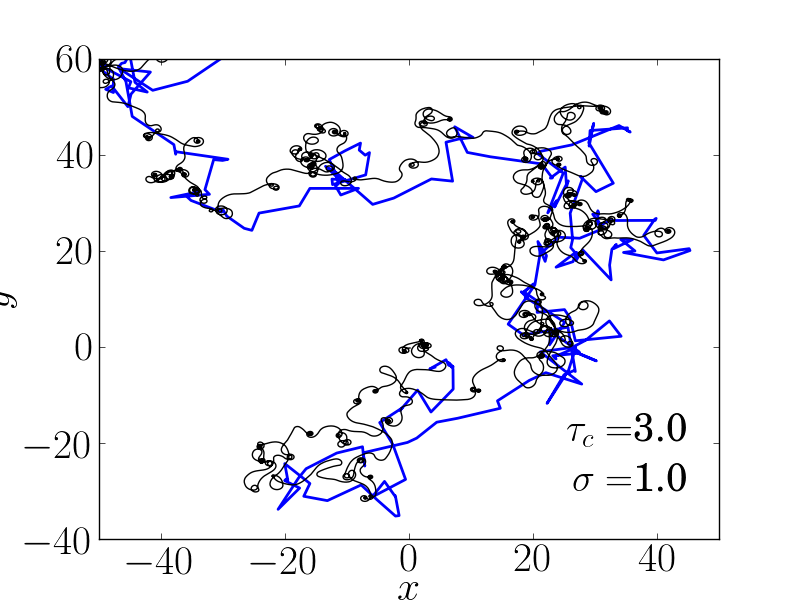}
  \includegraphics[width=.49\textwidth]{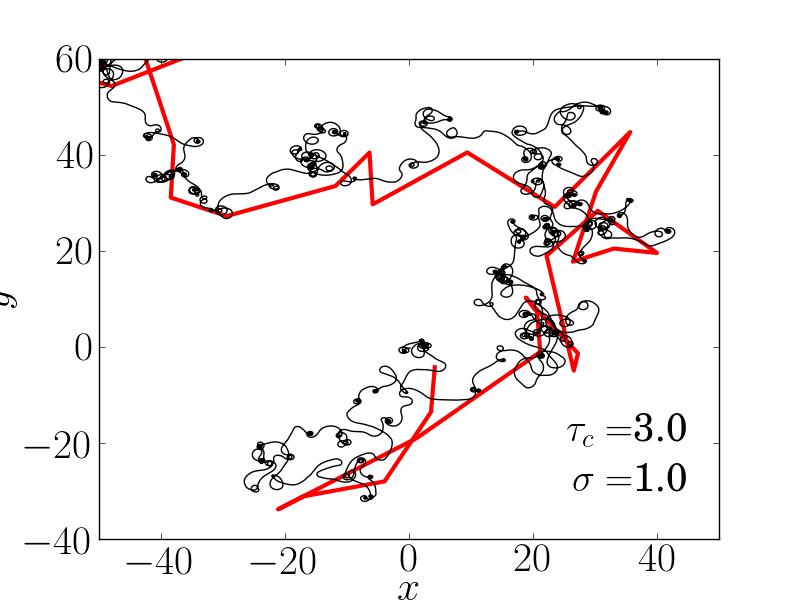}
  \caption{\small Sample trajectory of OUP driven by Cauchy noise, left: sampling time: $\Delta t=0.05s$ (black) and $\Delta t=5.0s$ (blue), right: $\Delta t=0.05s$ (black) and $\Delta t=40s$ (red), 
  colored trajectories shifted for better visibility }
  \label{figoup}
\end{figure}

At the end of this section we show a typical trajectory of the process and discuss the importance of sampling time lags.
Fig.(\ref{figoup}) shows always the same sample trajectory for $\sigma=1$ and $\tau_c=3.0$ with different 
sampling times (black: the sample time lag $\Delta t=0.05s$, blue: $\Delta t=5.0s$, red: $\Delta t=40s$). For the black 
trajectory positions every $\Delta t=0.05s$ where plotted and connected by a straight line. For the other sampling time lags
the same procedure was done accordingly.
For the small sampling time lag (black line) one sees a smooth structure with spirals or curls 
reminding of a run and tumble motion. The curly structure is influenced by the correlation time $\tau_c$. 
Higher correlation time $\tau_c$ increases length and size of a spiral. 
Setting the correlation time close to zero $\tau_c\approx 0$ removes the curls and makes the trajectory less smooth (not shown). 
The blue trajectory in Fig.(\ref{figoup}) is the same trajectory as the black one but sampled differently ($\Delta t=5.0s$); the sampling
time lag $\Delta t=5.0s$ still belongs to the lags at which non-Gaussian displacement distributions are observed. 
It still contains some information about the curls. Ignoring the underlying curly structure one might interpret the blue trajectory as one where 
a particle spends some time in a caged area or is rather immobile (maybe reorients itself) and then moves in rather straight stretches.
The red trajectory corresponds to sampling 
where the Gaussian displacement distribution of displacements during the time lags is established. 
There are no remnants of the spiral structure visible.

\section{Conclusion} 

We studied the stochastic dynamics of active particles moving at a constant
speed whose direction of motion is influenced by an
$\alpha$-stable noise source and by a constant torque. 
First, the noise was considered to be white, later on, we looked at a
colored Cauchy noise as generated by an Ornstein-Uhlenbeck process with the characteristic time $\tau_c$. 
The model with white noise generates the motion showing a crossover from the ballistic to
the diffusion behavior similar to a behavior for the Gaussian case. The behavior under the colored noise 
generates the pattern of motion strongly resembling the run and tumble situations 
or the behavior seen in experiments like \cite{Gunther} where particles 
spend some time in confined motion, reorient themselves and then move on.
For both physical models we derived analytical
expressions for the mean squared displacement and for the effective
diffusion coefficient. Astonishingly, in the case of Cauchy-noise, the existence of
correlations in the noise does not influence the MSD, the effective diffusion coefficient 
and the crossover time $\tau_D$ from ballistic to diffusive motion
coincide for both models  with $\alpha=1$. The distribution function of the 
displacements is however strongly influenced by correlations in the noise even at the 
times well inside the diffusive regime.  In particular, results of numeric
simulations for times larger $\tau_D$ can be well fitted by an
exponential distribution. This exponential behavior appears to be a transient.  A
crossover to Gaussian distribution of displacements takes place at a third
characteristic time $\tau_G > \tau_D$, connected with $\tau_c$ and $\tau_D$.
This colored noise mechanism might offer an approach to describe
observations of a transient non-Gaussian displacement distribution in
diffusion experiments like those performed by Wang et al.
\cite{Wang,Wangnat}.  The observed effect is absent in the case of the OUP 
with Gaussian white noise source, i.e with $\alpha=2$, and becomes the more
pronounced, the smaller is the value of $\alpha$.

Generally, our findings underline the importance to investigate the behavior
of the higher moments of displacement both in
experiments like \cite{Seuront,Bodecker,Edwards} and in simulations. These
moments might provide new information on the persistence of the motion. 
The studied case of the OUP with Cauchy noise is
special, and the independence of MSD of the correlation time $\tau_c$ of the noise is 
not a general situation. Nevertheless, we expect
deviations from a Gaussian distribution generally in models with a
correlated non-Gaussian noise in the angular dynamics, like in \cite{Weber_Sok}.

\section{Acknowledgments}
This work was developed within the scope of the IRTG 1740 funded by
DFG/FAPESP.  Lutz Schimansky-Geier acknowledges support from the
Humboldt-University at Berlin within the framework of excellence
initiative (DFG).  The authors thank Dr. B. Dybiec (Krakow) for
fruitful discussions.

\end{document}